\documentclass[a4paper,12pt]{article}
\usepackage{latexsym}
\usepackage{subeqnarray}
\usepackage{graphicx}

\makeatletter

\newcommand{\VCKM}{\ensuremath{V_\mathrm{CKM}}}
\newcommand{\UPMNS}{\ensuremath{U_\mathrm{PMNS}}}
\newcommand{\SM}{\mathrm{SM}}
\newcommand{\im}{\mathop{\operator@font Im}\nolimits} 
\newcommand{\diag}{\mathop{\operator@font diag}\nolimits} 
\newcommand{\PL}{\mathbf{L}}
\newcommand{\Slash}[1]{{\ooalign{\hfill$/$\hfill\crcr\hfill$#1$\hfill}}}

\makeatother

\begin{document}

\begin{titlepage}

\begin{flushright}
\begin{tabular}{l}
KEK-TH-1278
\end{tabular}
\end{flushright}

\vspace*{1cm}

\begin{center}
\Large
Ultraviolet divergences of flavor changing amplitudes
in the littlest Higgs model with T-parity
\end{center}

\begin{center}
\large
Toru Goto$^{a,}$\footnote{e-mail: {\tt tgoto@post.kek.jp}},
Yasuhiro Okada$^{a,b,}$\footnote{e-mail: {\tt yasuhiro.okada@kek.jp}}
and
Yasuhiro Yamamoto$^{a,b,}$\footnote{e-mail: {\tt yamayasu@post.kek.jp}}
\end{center}

\begin{center}
\it
$^a$Theory Group, KEK, Tsukuba, Ibaraki 305-0801, Japan
\\
$^b$Department of Particle and Nuclear Physics, The Graduate
University for Advanced Studies (Sokendai), Tsukuba, Ibaraki 305-0801,
Japan
\end{center}

\begin{center}
September 27, 2008
\end{center}

\begin{abstract}
Flavor changing neutral current processes are studied in the littlest
Higgs model with T-parity.
It is found that the logarithmic divergence reported earlier in $Z$
boson flavor changing processes is exactly canceled by contributions
from additional interaction terms of heavy fermions and the $Z$
boson.
Phenomenological impact on the $K\to\pi\,\nu\,\bar{\nu}$ processes is
discussed.  
\end{abstract}

\end{titlepage}

The little Higgs model \cite{ArkaniHamed:2001nc,ArkaniHamed:2002qy} was
proposed as a solution to the little hierarchy problem
\cite{Barbieri:1999tm} by canceling the quadratic divergence of the
Higgs mass term at one-loop level due to new diagrams with additional
gauge bosons and a heavy top-quark partner.
It was soon realized that the scale of the new particles should be in
the multi-TeV range in order to satisfy constraints from electroweak
precision measurements, so that the little hierarchy problem is
reintroduced \cite{Csaki:2002qg}.
This problem is avoided in the little Higgs model with T-parity
\cite{Cheng:2003ju}, because dangerous diagrams with the tree level
exchange of heavy neutral gauge bosons are forbidden by a new $Z_2$
discrete symmetry.

There are new sources of flavor transition in the littlest Higgs 
model with T-parity.
In order to assign the $Z_2$ symmetry for quarks and leptons, new heavy
fermions have to be introduced, and flavor mixing matrices associated
with the heavy fermions are independent of the
Cabibbo-Kobayashi-Maskawa (CKM) \cite{Cabibbo:1963yz} or the
Pontecorvo-Maki-Nakagawa-Sakata (PMNS) \cite{ref:PMNS} matrices
\cite{Hubisz:2005bd}.
Loop diagrams with heavy fermions and gauge bosons
can induce new contributions to quark flavor changing neutral 
current (FCNC) processes and lepton flavor violating (LFV) processes. 
In Refs.~\cite{Blanke:2006sb,Blanke:2006eb,Blanke:2007db}, these
processes are studied in detail and large deviation from the Standard
Model  predictions are shown to be possible.
In these works, it is argued that there are ``leftover'' singularities 
in FCNC/LFV amplitudes, and these logarithmic divergent terms 
can be dominant contributions.

In this letter we reconsider the FCNC precesses in the littlest Higgs
model with T-parity.
We evaluate $d^j\to d^i\,\nu\,\bar{\nu}$ ($i\neq j$) amplitude and find
disagreements with the previous results.
In particular, we point out that there are no leftover singularities in
the amplitude because the logarithmic divergent term in the $Z$-penguin
and box diagrams is exactly canceled by contributions from additional
interaction terms of heavy fermions and the $Z$ boson.
As a result, the FCNC amplitudes are determined by parameters
in the original Lagrangian, and do not have to depend on 
unknown parameters of the ultraviolet completion of the theory.

The model we study in this letter is the littlest Higgs model with
T-parity described in detail in Ref.~\cite{Blanke:2006eb}.
The gauge and the Higgs sector of the model is based on the
$SU(5)/SO(5)$ nonlinear sigma model.
The $SU(5)$ global symmetry is spontaneously broken down to $SO(5)$ by
the scalar field $\Sigma$, which is transformed as $\mathbf{15}$
representation of the $SU(5)$.
$\Sigma$ is written as a $5\times 5$ matrix and its $SU(5)$
transformation property is given as $\Sigma\to V \Sigma V^T$, where $V$
is an arbitrary $5\times 5$ unitary matrix with $\det V=1$.
The Nambu-Goldstone bosons are contained in a $5\times 5$ matrix $\xi$,
which is transformed under the $SU(5)$ as $\xi \to V \xi U$, where
$U=U(V,\xi)$ is an $SO(5)$ matrix determined by $V$ and $\xi$.
$[SU(2)\times U(1)]^2$ subgroup of the $SU(5)$ is
gauged and the $SU(2)\times U(1)$ electroweak gauge group of the
Standard Model is assumed to be a diagonal subgroup of the
$[SU(2)\times U(1)]^2$.

The vacuum expectation value of $\Sigma$ is chosen as
\begin{equation}
  \langle \Sigma \rangle = \xi_v \Sigma_0 \xi_v^T,
\qquad
  \xi_v = \langle \xi \rangle,
\end{equation}
\begin{equation}
  \Sigma_0 =
  \left(
    \begin{array}{ccccc}
      0 & 0 & 0 & 1 & 0 \\
      0 & 0 & 0 & 0 & 1 \\
      0 & 0 & 1 & 0 & 0 \\
      1 & 0 & 0 & 0 & 0 \\
      0 & 1 & 0 & 0 & 0
    \end{array}
  \right),
\end{equation}
\begin{equation}
  \xi_v =
\left(
  \begin{array}{ccccc}
    1 & 0 & 0 & 0 & 0 \\
    0 & \frac{c_{v}+1}{2} & \frac{is_{v}}{\sqrt{2}} & 0 & \frac{c_{v}-1}{2} \\
    0 & \frac{is_{v}}{\sqrt{2}} & c_{v} & 0 & \frac{is_{v}}{\sqrt{2}} \\
    0 & 0 & 0 & 1 & 0 \\
    0 & \frac{c_{v}-1}{2} & \frac{is_{v}}{\sqrt{2}} & 0 & \frac{c_{v}+1}{2}
  \end{array}
\right),
\label{eq:xi_v}
\end{equation}
where
$c_{v} = \cos(v/(\sqrt{2}f))$ and $s_{v} = \sin(v/(\sqrt{2}f))$.
$f=O(1\,\mathrm{TeV})$ breaks $SU(5)$ down to $SO(5)$ and
$[SU(2)\times U(1)]^2$ to $[SU(2)\times U(1)]_{\mathrm{SM}}$.
$v=(\sqrt{2}G_F)^{-1/2}$ is the electroweak symmetry breaking vacuum
expectation value.

The gauge bosons of $SU(2)_i\times U(1)_i$ ($i=1,2$) are denoted as
$W_i^{\pm,3}$ and $B_i$, respectively.
T-odd combinations $W_H^{\pm,3} = (W_1^{\pm,3} - W_2^{\pm,3})/\sqrt{2}$
and $B_H = (B_1 - B_2)/\sqrt{2}$ receive masses of $O(f)$.
T-even combinations $W_L^{\pm,3} = (W_1^{\pm,3} + W_2^{\pm,3})/\sqrt{2}$
and $B_L = (B_1 + B_2)/\sqrt{2}$ are identified as the Standard Model
electroweak gauge bosons.
After the electroweak symmetry breaking, the gauge fields of the mass
eigenstates are given as
\begin{eqnarray}
  Z_L &=& W_L^3 \cos\theta_W - B_L \sin\theta_W,
\qquad
  A_L = W_L^3 \sin\theta_W + B_L \cos\theta_W,
\\
  Z_H &=& W_H^3 \cos\theta_H - B_H \sin\theta_H,
\qquad
  A_H = W_H^3 \sin\theta_H + B_H \cos\theta_H,
\end{eqnarray}
where $A_L$ is the massless photon.
$\theta_W$ is the Weinberg angle which is determined by the
$[SU(2)\times U(1)]_{\SM}$ gauge coupling constants $g$ and $g'$ as
$\sin\theta_W=g'/\sqrt{g^2 + g^{\prime 2}}$.
The mixing angle of the T-odd gauge bosons is given by
\begin{equation}
  \tan2\theta_H =
  -\frac{g g' c_v^2 s_v^2}{ g^2 - g^{\prime 2}/5
    - (g^2 - g^{\prime 2})c_v^2 s_v^2/2}.
\end{equation}
Gauge boson masses are given as
\begin{eqnarray}
  m_{W_L}^2 &=& \frac{g^2 f^2}{2} s_v^2,
\qquad
  m_{Z_L}^2 = \frac{m_{W_L}^2}{\cos^2\theta_W},
\\
  m_{W_H}^2 &=& g^2 f^2 \left(1 - \frac{s_v^2}{2} \right),
\\
   m_{Z_H}^2 &=&
   \frac{g^2 f^2}{c_H^2-s_H^2}
   \left[
     \left( 1 - \frac{c_v^2 s_v^2}{2} \right) c_H^2
     -
     \frac{g^{\prime 2}}{5g^2}
     \left( 1 - \frac{5}{2} c_v^2 s_v^2 \right) s_H^2
   \right]
\nonumber\\&=&
  m_{W_H}^2 + O(\frac{v^4}{f^2}),
\\
   m_{A_H}^2 &=&
   \frac{g^{\prime 2} f^2}{5(c_H^2-s_H^2)}
   \left[
     \left( 1 - \frac{5}{2} c_v^2 s_v^2 \right) c_H^2
     -
     \frac{5g^2}{g^{\prime 2}}
     \left( 1 - \frac{c_v^2 s_v^2}{2} \right) s_H^2
   \right]
\nonumber\\&=&
 \frac{g^{\prime 2}f^2}{5}
  \left( 1 - \frac{5v^2}{4f^2} +  O(\frac{v^4}{f^4}) \right),
\end{eqnarray}
where $s_H=\sin\theta_H$ and $c_H=\cos\theta_H$.

In order to introduce the Standard Model fermions,
the $SU(5)\supset [SU(2)\times U(1)]^2$ symmetry structure has to be
extended because the $U(1)$ charges in the $SU(5)$ are not suitable to
accommodate the hypercharge of the Standard Model fermions.
This fact was implicitly assumed in Ref.~\cite{ArkaniHamed:2002qy}.
A straightforward way is adding two $U(1)$ factors.
Therefore, we consider the symmetry structure
$SU(5)\times U(1)''_1\times U(1)''_2 \supset
 [SU(2)_1 \times U(1)'_1 \times U(1)''_1] \times
 [SU(2)_2 \times U(1)'_2 \times U(1)''_2]$ hereafter.
The $U(1)$ subgroups of the $SU(5)$ are renamed as $U(1)'_{1,2}$.

\begin{table}[tbp]
\begin{displaymath}
\begin{array}{|c|cccc|cc|}
\hline
  & SU(2)_1 & SU(2)_2 & Y'_1 & Y'_2 & Y''_1 & Y''_2
\\
\hline
  q_1 = \left(\begin{array}{c} u_1 \\ d_1 \end{array}\right)
  & \mathbf{2} & \mathbf{1} & -\frac{3}{10} & -\frac{2}{10}
  & \frac{1}{3} & \frac{1}{3}
\\
  q_2 = \left(\begin{array}{c} u_2 \\ d_2 \end{array}\right)
  & \mathbf{1} & \mathbf{2} & -\frac{2}{10} & -\frac{3}{10}
  & \frac{1}{3} & \frac{1}{3}
\\
\hline
  t'_1
  & \mathbf{1} & \mathbf{1} & \frac{2}{10} & -\frac{2}{10}
  & \frac{1}{3} & \frac{1}{3}
\\
  t'_2
  & \mathbf{1} & \mathbf{1} & -\frac{2}{10} & \frac{2}{10}
  & \frac{1}{3} & \frac{1}{3}
\\
\hline
  \ell_1 = \left(\begin{array}{c} \nu_1 \\ e_1 \end{array}\right)
  & \mathbf{2} & \mathbf{1} & -\frac{3}{10} & -\frac{2}{10}
  & 0 & 0
\\
  \ell_2 = \left(\begin{array}{c} \nu_2 \\ e_2 \end{array}\right)
  & \mathbf{1} & \mathbf{2} & -\frac{2}{10} & -\frac{3}{10}
  & 0 & 0
\\
\hline
\end{array}
\end{displaymath}
\caption{Quantum numbers of the left-handed fermion fields.
Generation indices are suppressed.}
\label{tab:LHT-fermions-left}
\end{table}

\begin{table}[tbp]
\begin{displaymath}
\begin{array}{|c|cccc|cc|}
\hline
  & SU(2)_1 & SU(2)_2 & Y'_1 & Y'_2 & Y''_1 & Y''_2
\\
\hline
  u_R
  & \mathbf{1} & \mathbf{1} & 0 & 0
  & \frac{1}{3} & \frac{1}{3}
\\
  d_R
  & \mathbf{1} & \mathbf{1} & 0 & 0
  & -\frac{1}{6} & -\frac{1}{6}
\\
  q_{HR} = \left(\begin{array}{c} u_{HR} \\ d_{HR} \end{array}\right)
  & \mbox{--} & \mbox{--} & \mbox{--} & \mbox{--}
  & \frac{1}{3} & \frac{1}{3}
\\
\hline
  t'_{1R}
  & \mathbf{1} & \mathbf{1} & \frac{2}{10} & -\frac{2}{10}
  & \frac{1}{3} & \frac{1}{3}
\\
  t'_{2R}
  & \mathbf{1} & \mathbf{1} & -\frac{2}{10} & \frac{2}{10}
  & \frac{1}{3} & \frac{1}{3}
\\
\hline
  \nu_R
  & \mathbf{1} & \mathbf{1} & 0 & 0
  & 0 & 0
\\
  e_R
  & \mathbf{1} & \mathbf{1} & 0 & 0
  & -\frac{1}{2} & -\frac{1}{2}
\\
  \ell_{HR} = \left(\begin{array}{c} \nu_{HR} \\ e_{HR} \end{array}\right)
  & \mbox{--} & \mbox{--} & \mbox{--} & \mbox{--}
  & 0 & 0
\\
\hline
\end{array}
\end{displaymath}
\caption{%
Quantum numbers of the right-handed fermions.
Generation indices are suppressed.
$q_{HR}$ and $\ell_{HR}$ transform nonlinearly under
$[SU(2)_1\times U(1)'_1]\times [SU(2)_2\times U(1)'_2]$.
}
\label{tab:LHT-fermions-right}
\end{table}

The fermion sector of the littlest Higgs model with T-parity consists of
three families of quark and lepton fields $q_{1,2}^k$, $\ell_{1,2}^k$,
$u_R^k$, $d_R^k$, $\nu_R^k$, $e_R^k$, $q_{HR}^k$ and $\ell_{HR}^k$ with
the generation index $k=1,\,2,\,3$, and one set of the ``top partner''
fermions, $t'_{1,2}$ and $t'_{1R,2R}$.
Quantum numbers of these fermion fields are summarized in
Tables~\ref{tab:LHT-fermions-left} and \ref{tab:LHT-fermions-right}.
$Y'_{1,2}$ and $Y''_{1,2}$ are charges of $U(1)'_{1,2}$ and
$U(1)''_{1,2}$, respectively.
Subgroups $U(1)_i\subset U(1)'_i\times U(1)''_i$ ($i=1,2$) with the
charges $Y_i=Y'_i+Y''_i$ are gauged.
The Standard Model hypercharge $Y$ is given as $Y=Y_1+Y_2$.
``Mirror'' fermions $q_{HR}$ and $\ell_{HR}$ transform nonlinearly under
$[SU(2)_1\times U(1)'_1]\times [SU(2)_2\times U(1)'_2]$.
Transformation properties under T-parity are assigned as
$(q_1,\,t'_1,\,t'_{1R},\,\ell_1)\leftrightarrow
(-q_2,\,-t'_2,\,-t'_{2R},\,-\ell_2)$,
$(u_R,\,d_R,\,\nu_R,\,e_R)\leftrightarrow
(u_R,\,d_R,\,\nu_R,\,e_R)$ and
$(q_{HR},\,\ell_{HR}) \leftrightarrow (-q_{HR},\,-\ell_{HR})$.
From $q_i$, $t'_i$, $t'_{iR}$ and $\ell_i$ ($i=1,2$), T-even fermion
fields are constructed as
$q_L = (q_1 - q_2)/\sqrt{2}$,
$t'_+ = (t'_1 - t'_2)/\sqrt{2}$,
$t'_{+R} = (t'_{1R} - t'_{2R})/\sqrt{2}$ and
$\ell_L = (\ell_1 - \ell_2)/\sqrt{2}$,
while T-odd ones are given as
$q_H = (q_1 + q_2)/\sqrt{2}$,
$t'_- = (t'_1 + t'_2)/\sqrt{2}$,
$t'_{-R} = (t'_{1R} + t'_{2R})/\sqrt{2}$ and
$\ell_H = (\ell_1 + \ell_2)/\sqrt{2}$.
Note that $U(1)_{1,2}$ charges of the lepton doublets $\ell_{1,2}$ are
different from those used in Refs.~\cite{Hubisz:2004ft,Blanke:2006eb}.
This charge assignment enables us to treat
the lepton sector in the same way as the down-type quark sector.

$[SU(2)_1\times U(1)'_1\times U(1)''_1]\times
 [SU(2)_2\times U(1)'_2\times U(1)''_2]$ symmetric Yukawa coupling terms
of the down-type quarks are written as
\begin{equation}
{\cal L}_{\mathrm{down}} =
  \frac{i\lambda_d^{ij} f}{2\sqrt{2}}
  \sum_{p,q=1}^{2}\sum_{x,y,z=3}^{5}
  \epsilon_{pq} \epsilon_{xyz}
  \left[
    (\bar{\Psi}_2^{[\mathbf{\bar{5}}]i})_x
    (\Sigma)_{py} (\Sigma)_{qz}
    -
    (\bar{\Psi}_1^{[\mathbf{5}]i}\Sigma_0)_x
    (\tilde{\Sigma})_{py} (\tilde{\Sigma})_{qz}
  \right]
  d_R^j
  + \mathrm{H.c.},
\label{eq:Yukawa-down}
\end{equation}
where $\lambda_d^{ij}$ is the Yukawa coupling matrix and
$\tilde{\Sigma}$ is the T-parity image of $\Sigma$.
$\Psi_1^{[\mathbf{5}]i}$ and
$\Psi_2^{[\mathbf{\bar{5}}]i}$ are 
$SU(5)\times U(1)''_1\times U(1)''_2$ multiplets with quantum numbers
$(\mathbf{5},\,-\frac{1}{6},\,-\frac{1}{6})$ and
$(\mathbf{\bar{5}},\,-\frac{1}{6},\,-\frac{1}{6})$, respectively.
Quark doublets $q_{1,2}$ are embedded as
\begin{equation}
  \Psi_1^{[\mathbf{5}]} =
  \left(
    \begin{array}{c}
     q_1 \tilde{X}^\dagger \\ 0 \\ 0
    \end{array}
  \right),
\qquad
  \Psi_2^{[\mathbf{\bar{5}}]} =
  \left(
    \begin{array}{c}
      0 \\ 0 \\ q_2 X^\dagger
    \end{array}
  \right).
\label{eq:containersPsiprime}
\end{equation}
$X$ and $\tilde{X}$ are $SU(2)_i$ singlet scalar fields with the
$U(1)$ charges
$(Y'_1,\,Y'_2,\,Y''_1,\,Y''_2)=
(-\frac{4}{10},\, -\frac{6}{10},\, \frac{1}{2},\, \frac{1}{2})$ and
$(-\frac{6}{10},\, -\frac{4}{10},\, \frac{1}{2},\, \frac{1}{2})$,
respectively.
Following Ref.~\cite{Chen:2006cs}, we replace $X$ and $\tilde{X}$
with $(\Sigma_{33})^{-1/4}$ and its T-parity conjugate, respectively.
Since the $U(1)_1\times U(1)_2$ gauge charges of $X$ are the same as
those of $(\Sigma_{33})^{-1/4}$, this replacement maintains the gauge
invariance of (\ref{eq:Yukawa-down}), whereas the
$[U(1)'_1\times U(1)''_1]\times [U(1)'_2\times U(1)''_2]$ global
symmetry is explicitly broken.

After the diagonalization of mass matrices, we have the following T-even
(Dirac) fermions: three families of quarks and leptons $u^k$, $d^k$,
$\nu^k$ and $e^k$ ($k=1,\,2,\,3$), and one top partner quark $T_+$.
Similarly, T-odd fermions are $u_H^k$, $d_H^k$, $\nu_H^k$, $e_H^k$, and
$T_-$.
$T_+$ and all the T-odd fermions have masses of $O(f)$.
Flavor mixing among the T-even quarks is described by the CKM matrix
\VCKM\ and an extra angle in the mixing of the top quark $t=u^3$ and
$T_+$.
A parameter $x_L$ ($0<x_L<1$) is introduced for this extra mixing angle
\cite{Blanke:2006eb}.
The mixing matrix for the T-even lepton sector is the PMNS matrix
\UPMNS.
There are two independent mixing matrices in the T-odd gauge boson
coupling with the fermions.
We denote the mixing matrices at
$\bar{d}_H^i \Slash{Z}_H (1-\gamma_5) d^j$ and
$\bar{\nu}_H^i \Slash{Z}_H (1-\gamma_5) \nu^j$ couplings
as $(V_{Hd})_{ij}$ and $(V_{H\nu})_{ij}$, respectively.
Mixing matrices for other interaction terms are written in terms of
$V_{Hd}$, $V_{H\nu}$, \VCKM\ and \UPMNS.

Main difference between our results and those of
Ref.~\cite{Blanke:2006eb} originates from the gauge coupling of the
$Z_L$ boson and the ``mirror'' (right-handed and T-odd) fermions.
The kinetic and gauge interaction terms of the mirror
quarks are given as \cite{Hubisz:2004ft}
\begin{equation}
  {\cal L}_{\mathrm{kin}} =
  \frac{1}{2}
    \bar{\Psi}_R^{[\mathbf{5}]i}
    \gamma^\mu
    \left(
      i \partial_\mu
      +
      g \hat{W}_\mu
      +
      g' \hat{B}^{\Psi'_R}_\mu
    \right)
    \Psi_R^{[\mathbf{5}]i}
    +
    (\mbox{T-parity conjugate}).
\end{equation}
$\Psi_R^{[\mathbf{5}]i}$ is transformed as $\mathbf{5}$
representation under the $SU(5)$, and accommodate the mirror quarks as
\begin{equation}
  \Psi_R^{[\mathbf{5}]i} = \xi
  \left(
    \begin{array}{c}
      \tilde{\psi}_R^i \\ \chi_R^i \\ -i\sigma^2 q_{HR}^i
    \end{array}
  \right).
\end{equation}
$\tilde{\psi}_R^i$ and $\chi_R^i$ are assumed to decouple from the
effective theory.
The gauge fields are written as
\begin{eqnarray}
  \hat{W} &=&
  \frac{1}{2}
  \left(
    \begin{array}{ccccc}
      W_L^3 & \sqrt{2} W_L^+ & 0 & 0 & 0 \\
      \sqrt{2} W_L^- & -W_L^3 & 0 & 0 & 0 \\
      0 & 0 & 0 & 0 & 0 \\
      0 & 0 & 0 & -W_L^3 & -\sqrt{2} W_L^- \\
      0 & 0 & 0 & -\sqrt{2} W_L^+ & W_L^3
    \end{array}
  \right)
\nonumber\\&&
  +
  \frac{1}{2}
  \left(
    \begin{array}{ccccc}
      W_H^3 & \sqrt{2} W_H^+ & 0 & 0 & 0 \\
      \sqrt{2} W_H^- & -W_H^3 & 0 & 0 & 0 \\
      0 & 0 & 0 & 0 & 0 \\
      0 & 0 & 0 & W_H^3 & \sqrt{2} W_H^- \\
      0 & 0 & 0 & \sqrt{2} W_H^+ & -W_H^3
    \end{array}
  \right),
\\
  \hat{B}^{\Psi'_R} &=&
  \frac{1}{6}\diag( 7,\, 7,\, 4,\, 1,\, 1 )B_L
  +
  \frac{1}{10}\diag( 1,\, 1,\, -4,\, 1,\, 1 )B_H.
\end{eqnarray}
We obtain the gauge interaction term of $Z_L$ and $W_L$ bosons with the
mirror quarks as follows.
\begin{eqnarray}
  {\cal L}_{\mathrm{gauge}} &=&
 g_Z
 \left[
   \left( \frac{1}{2} - \frac{2s_W^2}{3} + \delta_v \right)
   \bar{u}_{HR}^i \Slash{Z}_L u_{HR}^i
   +
   \left( -\frac{1}{2} + \frac{s_W^2}{3} \right)
   \bar{d}_{HR}^i \Slash{Z}_L d_{HR}^i
 \right]
\nonumber\\&&
 +
 \frac{g}{\sqrt{2}}
 \left(1+\delta_v\right)
  \left(
    \bar{u}_{HR}^i \Slash{W}_L^+ d_{HR}^i
    +
    \bar{d}_{HR}^i \Slash{W}_L^- u_{HR}^i
  \right),
\label{eq:Lgauge-mirror}
\end{eqnarray}
where $g_Z=\sqrt{g^2+g^{\prime 2}}$, $s_W=\sin\theta_W$ and
$\delta_v = (c_v - 1)/2 = -v^2/(8f^2) + O(v^4/f^4)$.
The terms proportional to $\delta_v$ are missing in the Feynman
rules of Ref.~\cite{Blanke:2006eb}.
These terms are generated because of the fact that the gauge
field matrices $\hat{W}$ and $\hat{B}^{\Psi'_R}$ do not commute with
$\xi_v$.
The gauge interaction terms of the mirror leptons are derived in the
same way.
We find that there are the same correction terms that are proportional
to $\delta_v$ in the interactions of the mirror neutrinos $\nu_{HR}^i$.

\begin{figure}[tbp]
\centering
\begin{tabular}{cc}
\includegraphics[]{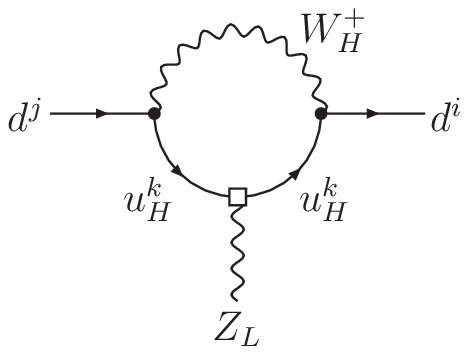} &
\includegraphics[]{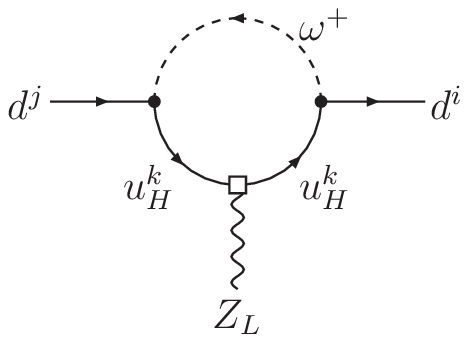} \\
 (a) & (b)
\end{tabular}
\caption{%
Feynman diagrams for additional contribution to the $Z_L$ penguin
amplitudes in $d^j\to d^i$ transition.
The box at the $Z_L$ vertex denotes the correction term which is
proportional to $\delta_v$.
$\omega^+$ in (b) is the Nambu-Goldstone boson absorbed by $W_H^+$.
}
\label{fig:lht-vZL}
\end{figure}

The additional interaction terms that are proportional to $\delta_v$ in
(\ref{eq:Lgauge-mirror}) affect the $Z_L$ penguin contributions to the
amplitudes for the $d^j\to d^i\,\nu\,\bar{\nu}$ FCNC processes.
Relevant Feynman diagrams are shown in Fig.~\ref{fig:lht-vZL}.
We calculate all the relevant $Z_L$ penguin and box diagrams with use of
the 't Hooft--Feynman gauge and obtain the Wilson coefficients for
$d^j\to d^i\,\nu\,\bar{\nu}$ as follows.
\begin{eqnarray}
  {\cal L}^{\mathrm{eff}}_{[d\nu]}
&=&
C_{[d\nu]LL}^{ijlm}
  (\bar{d}^i \gamma^\mu \PL d^j)(\bar{\nu}^l \gamma_\mu \PL \nu^m)
\,,
\\
C_{[d\nu]LL}^{ijlm}
&=&
  - \frac{g^4}{(4\pi)^2 m_{W_L}^2}
  \left[
    \delta_{lm}
    \left(
      \sum_k \lambda_k X_{\SM}(x_k)
      + \lambda_t \bar{X}_{\mathrm{even}}
    \right)
    +
    \sum_{k,n}
    \lambda^{H\nu}_n
    \xi_k
  J^{\nu\bar{\nu}}(z_k,y_n)
\right],
\nonumber\\
\label{eq:WCdnu}
\\
  J^{\nu\bar{\nu}}(z_k,y_n)
&=&
  \frac{v^2}{64 f^2}
\Biggl[
  2 z_k
  + z_k \log z_k
  + 3 z_k \mathbf{f}_{0[1]}(z_k)
  - \mathbf{f}_{3}(z_k,y_n)
  + 7 \mathbf{f}_{2}(z_k,y_n)
\nonumber\\&&\hphantom{\frac{v^2}{64 f^2}}
  -
  12 \mathbf{g}_{2[1]}(z_k,y_n)
  -
  \frac{3 r}{25}
  \mathbf{g}_{2[1]}(\frac{z_k}{r},\frac{y_n}{r})
  +
  \frac{6 r}{5}
  \mathbf{g}_{2}(z_k,y_n,r)
\Biggr],
\label{eq:Jnunu}
\end{eqnarray}
where
$\PL = \frac{1}{2}(1-\gamma_5)$,
$x_k = m_{u^k}^2/m_{W_L}^2$,
$z_k = m_{u_H^k}^2/m_{W_H}^2$,
$y_n = m_{e_H^n}^2/m_{W_H}^2$,
$r = m_{A_H}^2/m_{Z_H}^2$,
$\lambda_k = (\VCKM^*)_{ki} (\VCKM)_{kj}$,
$\lambda^{H\nu}_n = (V_{H\nu}^*)_{nl} (V_{H\nu})_{nm}$,
and
$\xi_k = (V_{Hd}^*)_{ki} (V_{Hd})_{kj}$.
Higher order terms in $v/f$ expansions are neglected.
We define the following classes of loop functions which are used in
(\ref{eq:Jnunu}).
\begin{subeqnarray}
  \mathbf{f}_{n}(y,z) &=& \frac{1}{y-z}
  \left( \frac{ y^n \log y }{y-1} - \frac{ z^n \log z }{z-1}\right),
\\
  \mathbf{g}_{n}(x,y,z) &=&
  \frac{\mathbf{f}_n(x,z)-\mathbf{f}_n(y,z)}{x-y},
\\
  \mathbf{f}_{n[1]}(z) &=& \mathbf{f}_n(z,1)
= \frac{ z^n \log z }{(z-1)^2} - \frac{1}{z-1},
\\
  \mathbf{g}_{n[1]}(x,y) &=& \mathbf{g}_{n}(x,y,1)
=
  \frac{\mathbf{f}_{n[1]}(x)-\mathbf{f}_{n[1]}(y)}{x-y}.
\end{subeqnarray}
For the T-even components of the Wilson coefficient (\ref{eq:WCdnu}),
namely $X_{\SM}$ and $\bar{X}_{\mathrm{even}}$, we obtain the same
results as those in Ref.~\cite{Blanke:2006eb}.
However, our result of the T-odd component, Eq.~(\ref{eq:Jnunu}),
differs from the corresponding formula of Ref.~\cite{Blanke:2006eb} in
the following two points.
\begin{itemize}
\item
There is no ``leftover'' $1/\epsilon$ singularity mentioned in
\cite{Blanke:2006eb}.
The term proportional to $\delta_v$ in the
$\bar{u}_{HR} \Slash{Z}_L u_{HR}$ coupling induces the following
contribution to $J^{\nu\bar{\nu}}$ through the diagrams shown in
Fig.~\ref{fig:lht-vZL}.
\begin{equation}
  \delta J^{\nu\bar{\nu}} =
  -\frac{v^2}{64f^2}
  z_k
  \left[
     \frac{1}{\epsilon} - \log\frac{m_{W_H}^2}{\mu^2}
     - \frac{1}{2} - \log z_k - 3 \mathbf{f}_{0[1]}(z_k)
  \right].
\label{eq:deltaJnunu}
\end{equation}
The $1/\epsilon$ term in (\ref{eq:deltaJnunu}) cancels the
singularities from other contributions.
For a crosscheck, we calculate $\delta J^{\nu\bar{\nu}}$ in the
unitary gauge, with the diagram Fig.~\ref{fig:lht-vZL}(a) only, and
obtain the same result.
\item 
The sign of the last term in the square brackets is opposite.
This term is generated by the box diagrams which consists of $d_H^k$,
$\nu_H^n$, $Z_H$ and $A_H$ for the internal lines.
The sign change is a consequence of the gauge charge assignments for the
lepton doublets in Table~\ref{tab:LHT-fermions-left}.
\end{itemize}

In the same way, we obtain the corresponding quantity
$J^{\mu\bar{\mu}}$, which appears in the $d^j\to d^i\,\ell^+\,\ell^-$
Wilson coefficient, as follows.
\begin{eqnarray}
  J^{\mu\bar{\mu}}(z_k,y_n) &=&
  \frac{v^2}{64 f^2}
\Biggl[
     2 z_k + z_k \log z_k + 3 z_k \mathbf{f}_{0[1]}(z_k)
  - \mathbf{f}_{3}(z_k,y_n)
 + 7 \mathbf{f}_{2}(z_k,y_n)
\nonumber\\&&\hphantom{\frac{v^2}{64 f^2}}
 + 6 \mathbf{g}_{2[1]}(z_k,y_n)
    +
  \frac{3 r}{25}
  \mathbf{g}_{2[1]}(\frac{z_k}{r},\frac{y_n}{r})
    +
  \frac{6 r}{5}
  \mathbf{g}_{2}(z_k,y_n,r)
\Biggr].
\label{eq:Jmumu}
\end{eqnarray}
The difference from the formula in Ref.~\cite{Blanke:2006eb} is similar
to the case of $J^{\nu\bar{\nu}}$.

In Ref.~\cite{Blanke:2007db}, the lepton flavor violating processes are
studied.
The formulae of the Wilson coefficients for the LFV effective operators
are derived in a similar manner to those for the quark FCNC's in
Ref.~\cite{Blanke:2006eb}, so that the results depend on the ultraviolet
singularities.
However, these singularities are also canceled by the $Z_L$ penguin
contributions induced by the $\delta_v$ terms in the $Z_L$ coupling with
mirror neutrinos.

\begin{figure}[tbp]
\centering
\includegraphics[scale=0.5]{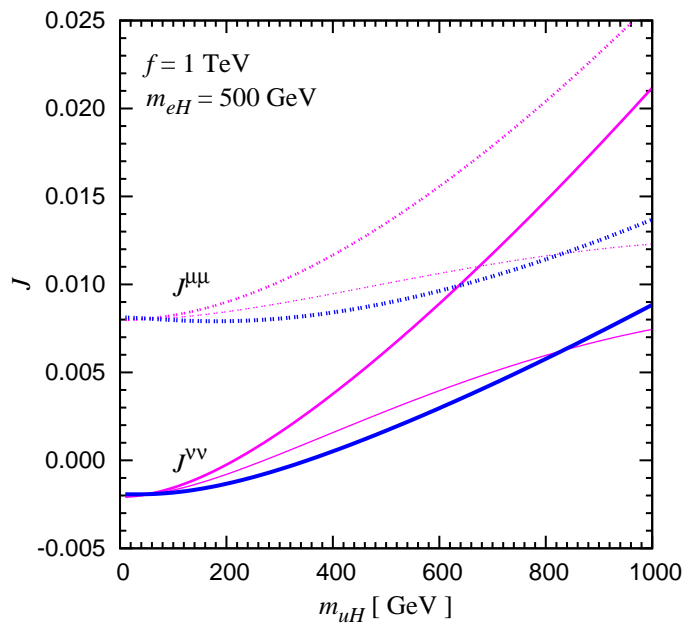}
\caption{%
$J^{\nu\bar{\nu}}$ (solid lines) and $J^{\mu\bar{\mu}}$ (dashed lines)
as functions of the mirror quark mass $m_{uH}$.
Dark-grey/blue lines show the values calculated with Eqs.~(\ref{eq:Jnunu})
and (\ref{eq:Jmumu}).
Light-grey/magenta lines are the results of 
Ref.~\cite{Blanke:2006eb} with $J^{\nu\bar{\nu}}_{\mathrm{div}}
= J^{\mu\bar{\mu}}_{\mathrm{div}}
= \frac{v^2}{64f^2}z\log\frac{(4\pi f)^2}{m_{W_H}^2}$ (thick lines) and
$J^{\nu\bar{\nu}}_{\mathrm{div}}=J^{\mu\bar{\mu}}_{\mathrm{div}}=0$
(thin lines).
}
\label{fig:JnunuJmumu}
\end{figure}

In Fig.~\ref{fig:JnunuJmumu}, we show numerical values of
$J^{\nu\bar{\nu}}$ and $J^{\mu\bar{\mu}}$ evaluated with use of
Eq.~(\ref{eq:Jnunu}) and (\ref{eq:Jmumu}), respectively.
We take the mirror lepton mass $m_{eH}=500\,\mathrm{GeV}$ and
$f=1\,\mathrm{TeV}$ for the calculation.
We also show the values obtained by the formulae given in
Ref.~\cite{Blanke:2006eb} (with $J^{\nu\bar{\nu}}_{\mathrm{div}}
= J^{\mu\bar{\mu}}_{\mathrm{div}}
= \frac{v^2}{64f^2}z\log\frac{(4\pi f)^2}{m_{W_H}^2}$ and
$J^{\nu\bar{\nu}}_{\mathrm{div}}= J^{\mu\bar{\mu}}_{\mathrm{div}}=0$) in
the same plot for comparison.
The results with Eq.~(\ref{eq:Jnunu}) and Eq.~(\ref{eq:Jmumu})
look close to the results Ref.~\cite{Blanke:2006eb} with
$J^{\nu\bar{\nu}}_{\mathrm{div}}=J^{\mu\bar{\mu}}_{\mathrm{div}}=0$.
Most of the numerical differences come from the contribution of
$\delta J^{\nu\bar{\nu}}$.
The effect of the sign change of the $d_H^k-\nu_H^n-Z_H-A_H$ box diagram
is small.

\begin{figure}[tbp]
\centering
\begin{tabular}{cc}
\includegraphics[scale=0.6]{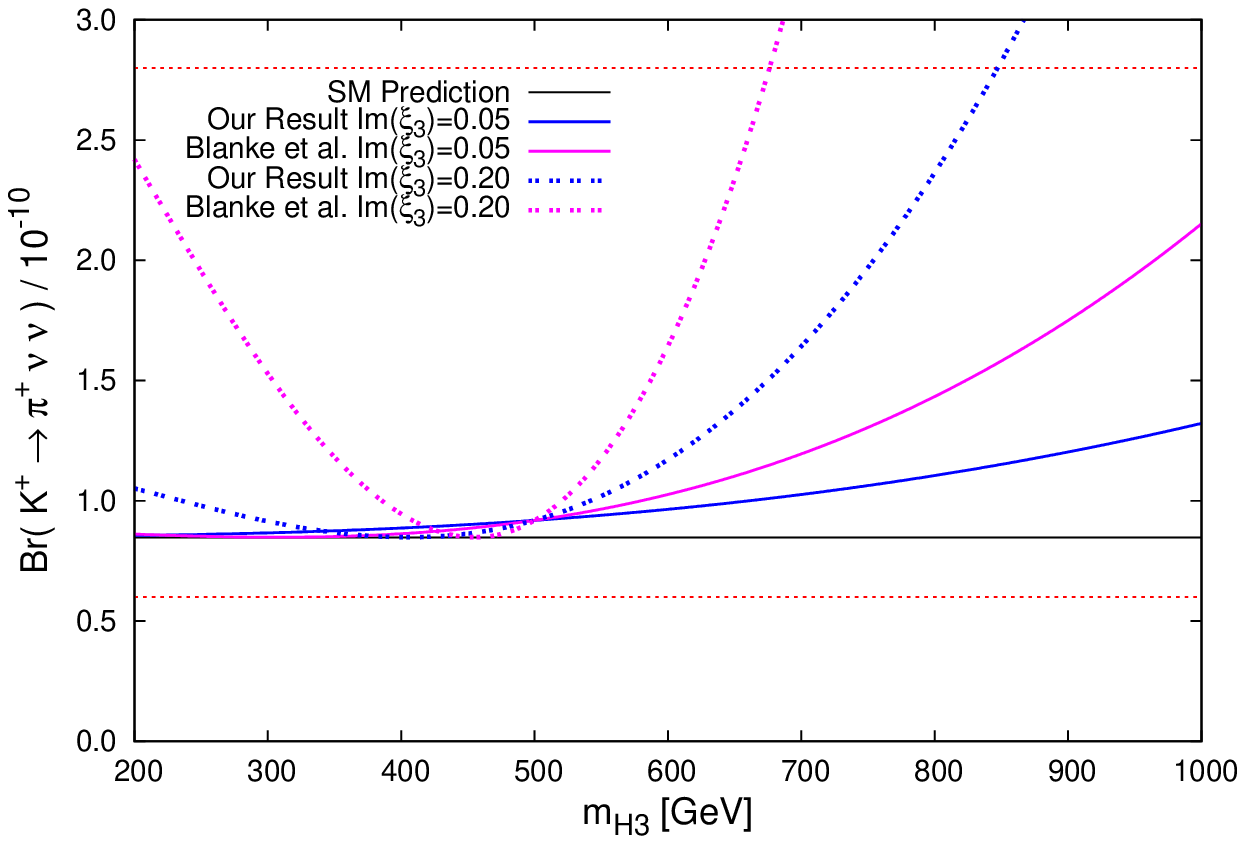} &
\includegraphics[scale=0.6]{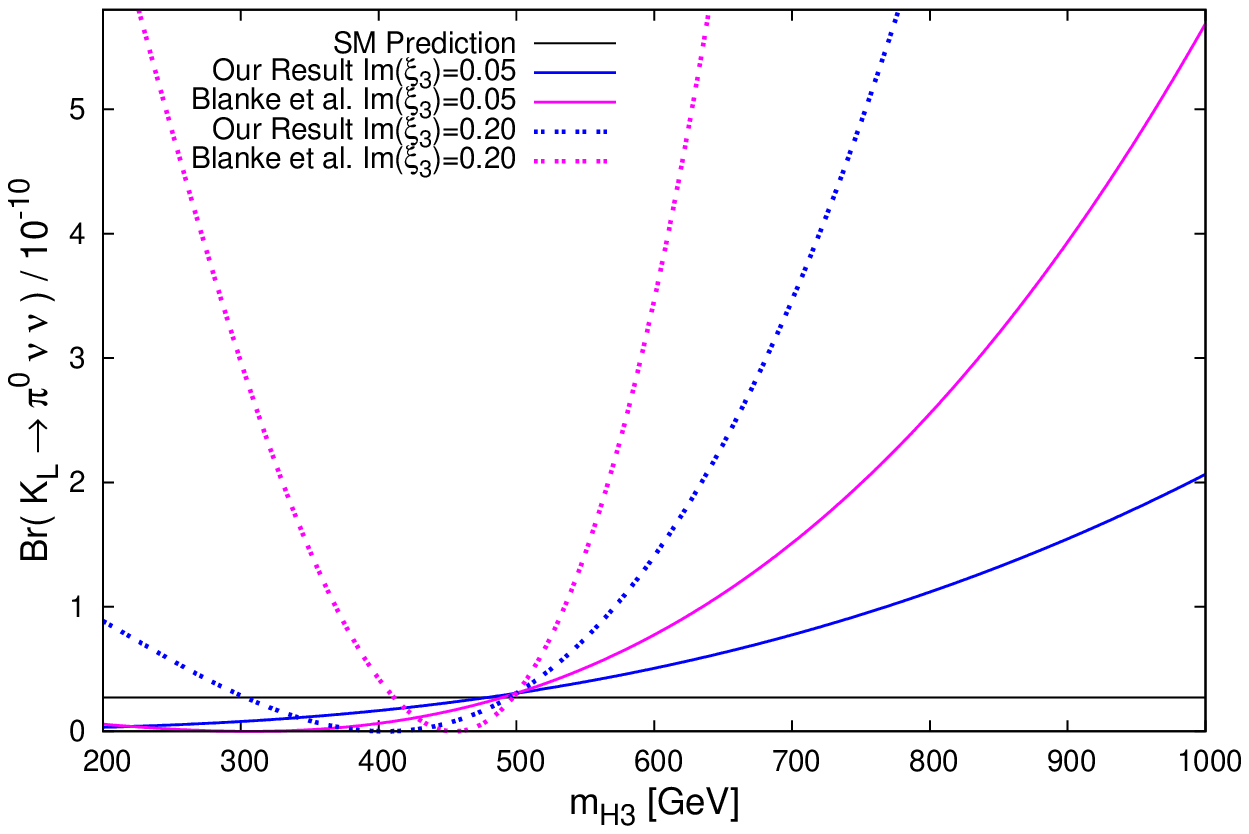} \\
(a) & (b)
\end{tabular}
\caption{%
Branching ratios of (a) $K^+\to\pi^+\,\nu\,\bar{\nu}$ and
(b) $K_L\to\pi^0\,\nu\,\bar{\nu}$ for $\im\xi_3=0.05$ (solid line) and
$\im\xi_3=0.20$ (dashed line).
Dark-grey/blue lines show the values calculated with
Eq.~(\ref{eq:Jnunu}).
Light-grey/magenta lines are the results of 
Ref.~\cite{Blanke:2006eb} with $J^{\nu\bar{\nu}}_{\mathrm{div}}
= \frac{v^2}{64f^2}z\log\frac{(4\pi f)^2}{m_{W_H}^2}$.
Horizontal solid lines show the Standard Model predictions.
Dotted horizontal lines in (a) show $1\sigma$ range of the experimental
value \cite{Anisimovsky:2004hr}.
}
\label{fig:kpnn}
\end{figure}

Our result affects the predictions for $\Delta F=1$ FCNC and LFV
processes, in which $Z_L$ penguin contributions of $O(v^2/f^2)$ are
relevant.
Here we present the results for the $K\to\pi\,\nu\,\bar{\nu}$ processes.
In Fig.~\ref{fig:kpnn}, the branching ratios of
$K^+\to\pi^+\,\nu\,\bar{\nu}$ and $K_L\to\pi^0\,\nu\,\bar{\nu}$ are
shown as functions of the third generation T-odd quark mass
$m_{H3}=m_{u_H^3}\simeq m_{d_H^3}$.
As for other input parameters, we adopt the ``Scenario 6'' in
Ref.~\cite{Blanke:2006eb}.
In this scenario, it is assumed that the T-odd quark masses of the first
and the second generations are equal and that the mixing parameter
$\xi_3^{(K)}=(V_{Hd}^*)_{3s} (V_{Hd})_{3d}$ is pure imaginary.
This parameter choice makes the T-odd contribution to the $K^0-\bar{K}^0$
mixing negligibly small.
We take the first and the second generation T-odd quark masses and all
the T-odd lepton masses as $500\,\mathrm{GeV}$.
The parameters $f$ and $x_L$ are fixed as $f=1\,\mathrm{TeV}$ and
$x_L=0.5$.
For comparison, we also plot the branching ratios calculated by the
formulae in Ref.~\cite{Blanke:2006eb} with
$J^{\nu\bar{\nu}}_{\mathrm{div}} =
\frac{v^2}{64f^2}z\log\frac{(4\pi f)^2}{m_{W_H}^2}$.
We see that both branching ratios can be significantly larger than the
Standard Model prediction, although our values of the T-odd
contributions are smaller than those in Ref.~\cite{Blanke:2006eb} in
general.

In conclusion, we have revisited FCNC processes in the littlest Higgs
model with T-parity.
We have found that there is no divergence in the
$d^j\to d^i\,\nu\,\bar{\nu}$ amplitude, that was reported earlier.
This implies that FCNC processes can be insensitive to the physics
at the cut-off scale.
This is in contrast with the case of the littlest Higgs model without
T-parity \cite{Buras:2006wk}, where the leftover singularity exists.
We have also re-evaluated the branching ratios of
$K\to\pi\,\nu\,\bar{\nu}$ decays and found significant changes from
previous results.
We expect similar changes in other FCNC and LFV observables.

The work of T.\ G.\ and Y.\ O.\ is supported in part by the Grant-in-Aid for
Science Research, Ministry of Education, Culture, Sports, Science and
Technology, Japan, No.\ 16081211 and by the  Grant-in-Aid for Science
Research, Japan Society for the Promotion of Science, No.\ 20244037.

\end{document}